\documentstyle [preprint,prd,aps] {revtex}
\begin{document}
\draft
\title {Two-photon reactions with KLOE detector at DA$\Phi$NE}

\author { Saro Ong\footnote{e-mail : ong@ipno.in2p3.fr}}

\address{ Institut de Physique Nucl\'eaire, IN2P3-CNRS, Universit\'e
Paris-Sud, 91406 Orsay Cedex, France}

\maketitle
\vskip 2 true cm
\begin{center}
{\bf {Abstract} }
\end{center}

{\small We reexamine the feasibility of two-photon reactions at DA$\Phi$NE
with the KLOE detector excluding the small angle tagging system. Event-rate
predictions of interesting channels :
$\gamma \gamma \rightarrow \pi^0~,~\eta$ and
$\gamma \gamma \rightarrow \pi^+ \pi^-~, ~\pi^0 \pi^0$
are discussed.}     
  
\vskip 2 true cm

\section{  Introduction }

The physical interest of $\gamma \gamma$ collisions at DA$\Phi$NE 
has first been pointed
out by A. Courau [1]. He has shown that the photons remain quasi-real
up to a relatively large angle (a few hundreds of mrad.) of the scattered
electrons. The counting rate of a double-tag measurement [2] remains 
relatively
important (Fig. 1) in contrast with that of a similar measurement at 
a high energy machine (LEP) [3].\\

It has been shown that there is an overwhelming background
the $\phi$-meson production through annihilation process.
Therefore a study of
precision measurements of two-photon reactions
with the KLOE detector [4] using the small angle tagging system has been
performed [5]. For $\gamma \gamma \rightarrow \pi^0 \pi^0$,
a transverse-momentum cut allows one to isolate the signal from
the background due to
$\phi \rightarrow K_S K_L \rightarrow \pi^0 \pi^0~ +~ X $ (undetected).\\

The KLOE detector allows for a minimal tagging angle of the electrons
of about 200 mrad, which is a rather large angle. In this context, 
we reexamine
the feasibility of the two-photon reactions at DA$\Phi$NE with
this constraint. Realistic event-rate predictions of interesting
channels (pseudoscalar meson and pion-pair productions) are shown
in sections II and III.

\section{Pseudoscalar meson production}

For pseudoscalar meson production by $\gamma \gamma^*$
collisions, there has been considerable theoretical and experimental
activity to predict and measure the
pion-photon transition form factor $F_{\pi\gamma}(Q^2)$.
Previous pQCD predictions [6-9], using the asymptotic distribution amplitude,
have been found to be consistent with the CELLO
[10] and CLEO II data [11] up to $Q^2=8.$ GeV$^2$. The leading order (LO) 
prediction
of the pion-photon transition form factor, in the framework of the hard
scattering approach [12], reads :

\begin{equation}
\displaystyle{ F_{\pi\gamma}(Q^2)={2f_{\pi} \over Q^2}}
~~~~~~~~~~~~~~~~~~~~~~{\rm as} ~~~~~Q^2 \rightarrow \infty
\end{equation}
where $f_{\pi} \simeq 93$ MeV is the pion decay constant.\\

Interpolating between this asymptotic expression and the current-algebra
prediction at $Q^2 \rightarrow 0$, Brodsky and Lepage (BL) have proposed a
simple-pole formula [6] :
\begin{equation}
\displaystyle{ F_{\pi\gamma}(Q^2)={1 \over 4\pi^2 f_{\pi}}
~{1 \over (1+Q^2/\Lambda_{\pi}^2)}}
\end{equation}
with $\Lambda_{\pi}^2=8 \pi^2 f_{\pi}^2$; one gets, for the 
mass-scale parameter, $\Lambda_{\pi} \simeq 826$ MeV. However, this
formula is compatible as well with the prediction based on vector-meson
dominance model (VMD) with $\Lambda_{\pi}=m_{\rho}$.\\
CLEO Collaboration [11] has reported the pole mass fit of the pion-photon
transition form factor. They obtain $\Lambda_{\pi^0}=766\pm 10\pm 12\pm 16$
MeV, a value close to the mass of the $\rho$-meson.\\

Recently a full calculation, assuming the asymptotic distribution amplitude
and including a QCD radiative correction [13], has been performed [14]. 
Formula (1) is modified, taking the following form 
\begin{equation}
\displaystyle{ F_{\pi\gamma}(Q^2)={2f_{\pi} \over Q^2} \left (1-
{5 \alpha_{V}(e^{-3/2}Q) \over 3 \pi} \right )}
\end{equation}
Assuming $\alpha_{V}(e^{-3/2}Q)/\pi \simeq 0.12$, the magnitude
of $Q^2F_{\pi\gamma}(Q^2)$ is remarquably consistent with the CLEO data
(see Fig. 2). One can also fit the data by using the interpolation
formula (2) with
a new mass scale $\Lambda_{\pi} \simeq 739$ MeV. So it is very hard to
discriminate between the BL model and the VMD.\\

Let us notice that the slope predictions in the framework of
chiral perturbation theory (CHPT) [15] are consistent with the
pole mass fit of CLEO. 
It is very exciting
that with this simple process a nice description of the form factor is
reliable from $Q^2=0$ up to $Q^2 \rightarrow \infty$.\\

Kessler and Ong [16] have shown that pseudoscalar-meson production
by two off-shell photons can be used to check the pQCD
hard scattering approach. Notice that in this approach the transition 
form factor
is $\sim Q^{-2}$ while the VMD predicts it to be $\sim Q^{-4}$.
In the symmetric configuration $Q^{\prime 2}= Q^2$ where
$Q^2=-q^2~,~Q^{\prime 2}=-q^{\prime 2}~(q, q^{\prime}$ are the four-momenta
of the photons) the form factor becomes independent of the choice of the
distribution amplitude; one obtains with the BL interpolation formula 
\begin{equation}
\displaystyle{ F_{\pi\gamma^*}(Q^2,Q^2)={1 \over 4\pi^2 f_{\pi}}
~{1 \over (1+3Q^2/\Lambda_{\pi}^2)}}
\end{equation}
which is compared with the VMD prediction :
\begin{equation}
\displaystyle{ F_{\pi\gamma^*}(Q^2,Q^2)={1 \over 4\pi^2 f_{\pi}}
~{1 \over (1+Q^2/m_{\rho}^2)^2}}
\end{equation}

The measurement of this form factor at low $Q^2$ can be performed at
DA$\Phi$NE with the KLOE detector. In the framework of CHPT, the
slope prediction of $\pi^0 \rightarrow \gamma^*$ transition form factor 
is similar as well with the prediction based on
the vector-meson dominance (VMD).
The slope determination at
$Q^2 \rightarrow 0$
of $F_{\pi\gamma^*}(Q^2,Q^2)$ 
should allow one to check the validity of the BL model vs
the CHPT. 

In order to check the feasibility of such measurements, we have computed
the number of events for energy and luminosity planned at DA$\Phi$NE 
assuming the VMD form factor (see table 1).

\section{Pion pairs production}

In a previous paper [19], we have shown that azimuthal correlations in 
single-tag measurements of photon-photon collisions can be used to
check dynamic models We here extend our investigation to double-tag
measurements.\\

The main contribution to $\gamma \gamma \rightarrow \pi^+ \pi^-$
arises from the Born terms. The chiral loops give the
next order contribution [17] and are consistent with the MARK II data [18].
It had been shown [19-21] that azimuthal correlations can be used
to check dynamical models. At DA$\Phi$NE, with somewhat large angles
of electron tagging, it should be possible to determine these correlations
in double-tag measurements of photon-photon collisions.\\

In contrast with the charged-pion pair production, the process
$\gamma \gamma \rightarrow \pi^0 \pi^0$ involves no contribution
from Born terms. A finite one-loop contribution up to ${\cal O} (p^4)$
in the framework of CHPT has been computed [22]. Comparing with the presently
available data from Crystal Ball [23], this prediction lies below them
within 2$\sigma$. Recently, the amplitude involving two loops has 
been evaluated
[24]; the corresponding cross section prediction 
agrees rather well with the Crystal
Ball data. However, another prediction [25] based on the dispersion
relations has been found to be as well consistent with the data. 
We emphasize
the importance of precise measurements of the azimuthal correlations
at DA$\Phi$NE for this channel.\\

In double-tag measurements where both electrons are tagged at small angle
$(Q,~Q^{\prime} \ll W/2)$, we can use the 5-term formula [20,21]. Integrating
the differential cross section over all variables other than $\phi_1$
and $\phi_2$, we obtain :
\begin{equation}
\displaystyle{{d\sigma \over d\phi_1 d\phi_2}=\sigma_0+\sigma_1
\cos2\phi_1+\sigma_2\cos2\phi_2+\sigma_3 \cos2(\phi_1+\phi_2)+\sigma_4
\cos2(\phi_1-\phi_2)}
\end{equation}
where $\phi_1$ and $\phi_2$ are the azimuthal angles, in the 
$\gamma^* \gamma^*$
c.m. frame, between one of the particles (pions) produced and the two outgoing
electrons. The helicity terms $\sigma_0\ldots \sigma_4$ 
can be determined from measurements of the integrated 
cross section and azimuthal correlations.\\

For numerical predictions, we assume the general experimental
conditions for the double-tag case (see tables 2-3), as it is more interesting
to study azimuthal correlations in double-tag measurements. One
obtains different predictions (figs. 3-4) for $\pi^+\pi^-$ and
$\pi^0 \pi^0$.\\

A complete and exact Monte Carlo for $\gamma \gamma \rightarrow \mu^+ \mu^-$
is also available [3]. It can be used to calibrate the measurement of these
azimuthal correlations.\\

In conclusion, we are showing that, for two-pion production, some useful
information may be provided by the study of azimuthal correlations. Also the
possibility of a sizeable two-loop effect in the framework of CHPT or
a one loop effect in GCHPT [26]
has been discussed.

\section{conclusion}

We have shown that DA$\Phi$NE is a unique $e^+e^-$ machine where a
double-tag measurements of $\gamma^* \gamma^*$ collisions should be performed.
It will be the first time that the $\pi^0 \rightarrow \gamma^*$ 
transition form
factor at low $Q^2$ and the five structure functions in pion pairs
production measurements should allow one to check dynamic models.

\acknowledgments
The author is grateful to A. Courau and P. Kessler for careful
reading of the manuscript and for useful discussions. The author acknowledge
partial support from the EEC-TMR Program, contract N. CT98-0169.

\begin{table}
\begin{center}
\caption{\small Number of events expected assuming $E_{beam}=0.51~$ GeV
with integrated luminosity $L=5.~10^{39}~cm^{-2}$. $Q_{min}^2$ is the
minimal value of the four-momentum squared of the virtual photon.}
\label{tab01}

\begin{tabular}{|c|c|c|}

$\gamma^{*} \gamma^{*} \rightarrow$ & $\pi^0$~~~~~~~~~~~~~~~~~~~ &
$\eta$~~~~~~~~~~~~~~~~~~~\\ \hline
$Q_{min}^2=0. $ GeV$^2$ &2.~$10^6$~~~~~~~~~~~~~~~~~~~&
5.~$10^5$~~~~~~~~~~~~~~~~~~~\\ \hline
$Q_{min}^2=5.~ 10^{-3}$ GeV$^2$ &3.55~$10^4$~~~~~~~~~~~~~~~~~~~&
9.18~$10^3$~~~~~~~~~~~~~~~~~~~\\ \hline
$Q_{min}^2=1.~ 10^{-2}$ GeV$^2$ &1.99~$10^4$~~~~~~~~~~~~~~~~~~~&
5.44~$10^3$~~~~~~~~~~~~~~~~~~~\\ \hline
$Q_{min}^2=5.~ 10^{-2}$ GeV$^2$ &2.63~$10^3$~~~~~~~~~~~~~~~~~~~ &
8.~$10^2$~~~~~~~~~~~~~~~~~~~\\
\end{tabular}
\end{center}
\end{table}

\begin{table}
\begin{center}
\caption{\small Number of events expected with $E_{beam}=0.51~$ GeV
and the integrated luminosity $L=5.~10^{39}~cm^{-2}$. We assume
the invariant mass
$ 2m_{\pi} \leq W_{\gamma\gamma} \leq 700$ MeV 
and an acceptance cut $|\cos\theta| \leq 0.8$}

\begin{tabular}{|c|c|c|}
$\gamma^{*} \gamma^{*} \rightarrow$ & $\pi^+ \pi^-$~~~~~~~~~~~~~~~~~~~ &
$\pi^0 \pi^0$~~~~~~~~~~~~~~~~~~~\\ \hline 
$0 \leq \theta_e \leq 300 $ mrad &9.54~ $10^5$~~~~~~~~~~~~~~~~~~~&
7.58~$10^3$~~~~~~~~~~~~~~~~~~~\\ \hline 
$200 \leq \theta_e \leq 300 $ mrad &~4.82~$10^3$~~~~~~~~~~~~~~~~~~~&
4.5~$10^1$~~~~~~~~~~~~~~~~~~~\\ \hline
$250  \leq \theta_e \leq 300 $ mrad & 9.25~$10^2$~~~~~~~~~~~~~~~~~~~ &
~$10$~~~~~~~~~~~~~~~~~~~\\
\end{tabular}
\end{center}

\label{tab02}
\end{table}

\begin{table}
\begin{center}
\caption{\small Same as Tab. II, but with $|\cos \theta| \leq 0.98$}

\begin{tabular}{|c|c|c|}
$\gamma^{*} \gamma^{*} \rightarrow$ & $\pi^+ \pi^-$~~~~~~~~~~~~~~~~~~~ &
$\pi^0 \pi^0$~~~~~~~~~~~~~~~~~~~\\ \hline
$0 \leq \theta_e \leq 300 $ mrad &2.11~ $10^6$~~~~~~~~~~~~~~~~~~~&
4.~$10^4$~~~~~~~~~~~~~~~~~~~\\ \hline 
$200 \leq \theta_e \leq 300 $ mrad &~7.93~$10^3$~~~~~~~~~~~~~~~~~~~&
2.~$10^2$~~~~~~~~~~~~~~~~~~~\\ \hline
$250  \leq \theta_e \leq 300 $ mrad & 1.45~$10^3$~~~~~~~~~~~~~~~~~~~ &
~$35$~~~~~~~~~~~~~~~~~~~\\
\end{tabular}
\end{center}

\label{tab03}
\end{table}

\begin{figure}[htbp]
\begin{center}
\leavevmode
\caption{Angular distribution of either scattered electron
for $e^+~e^- \rightarrow  e^+~e^-\pi^+ \pi^-$ with
$E_{beam}=510$ MeV and an acceptance cut $|\cos \theta| \leq 0.8$}
\label{fig:1}
\end{center}
\end{figure}
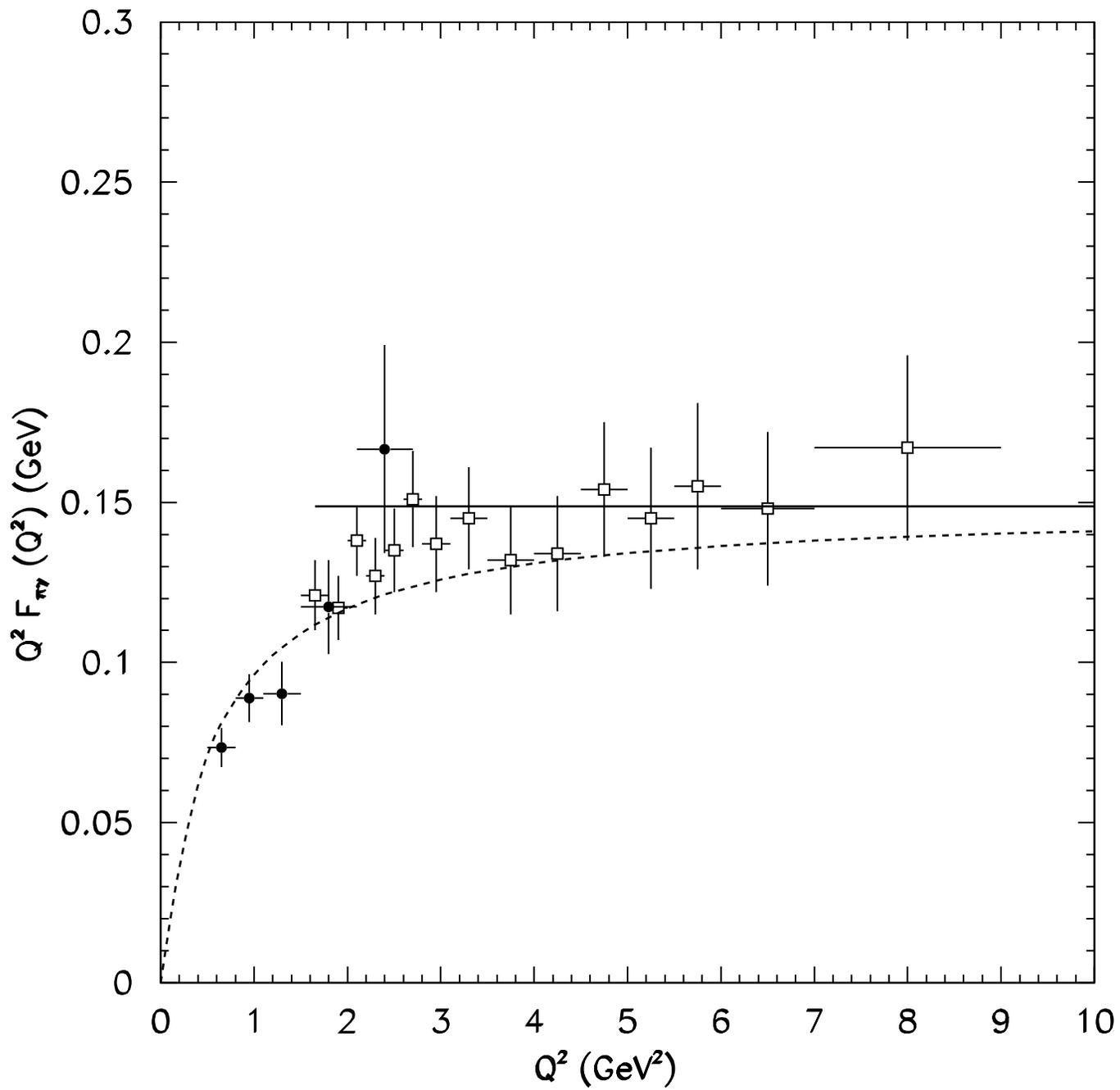
\begin{figure}[htbp]
\begin{center}
\leavevmode
\caption{ Solid line : prediction of the $\pi\gamma$ transition form factor, 
including the QCD radiative correction
and assuming the asymptotic distribution amplitude. Dash line : prediction
with the interpolation formula assuming $\Lambda_{\pi} \simeq 739$ MeV. 
Data are taken from Ref.[10,11]}
\label{fig:2}
\end{center}
\end{figure}
\begin{figure}[htbp]
\begin{center}
\leavevmode
\caption{Born-term predictions of azimuthal distributions with regard to  
($\phi_1$,$\phi_2$,~$\phi_1+\phi_2$
and $\phi_1-\phi_2$) for $e^+~e^- \rightarrow  e^+~e^-~\pi^+ \pi^-$ with
$E_{beam}=510$ MeV and an acceptance cut $|\cos \theta| \leq 0.8$}
\label{fig:3}
\end{center}
\end{figure}
\begin{figure}[htbp]
\begin{center}
\leavevmode
\caption{ Same as Fig. 3 for
$e^+~e^-  \rightarrow e^+~e^-~\pi^0 \pi^0$ in Chiral Perturbation Theory}
\label{fig:4}
\end{center}
\end{figure}

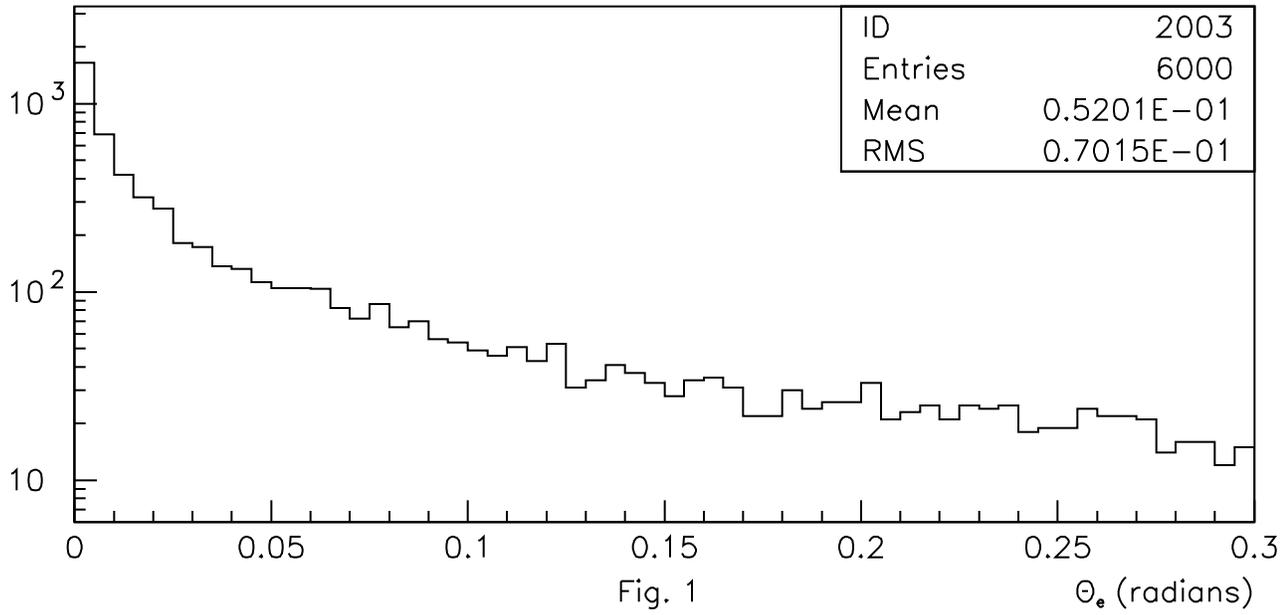
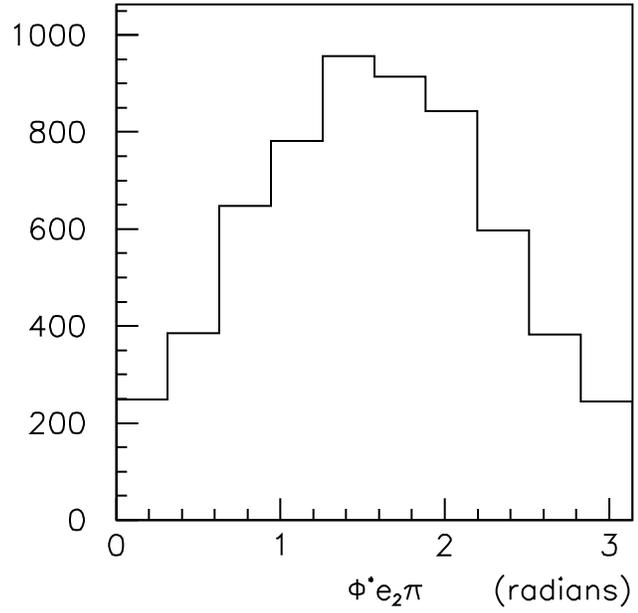
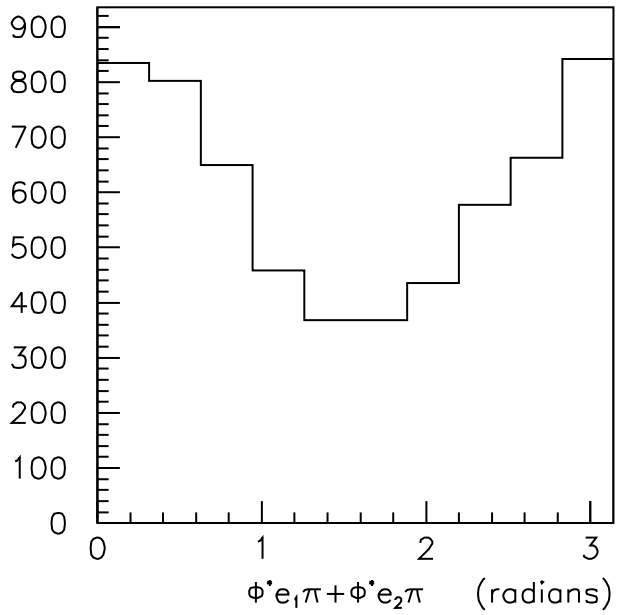
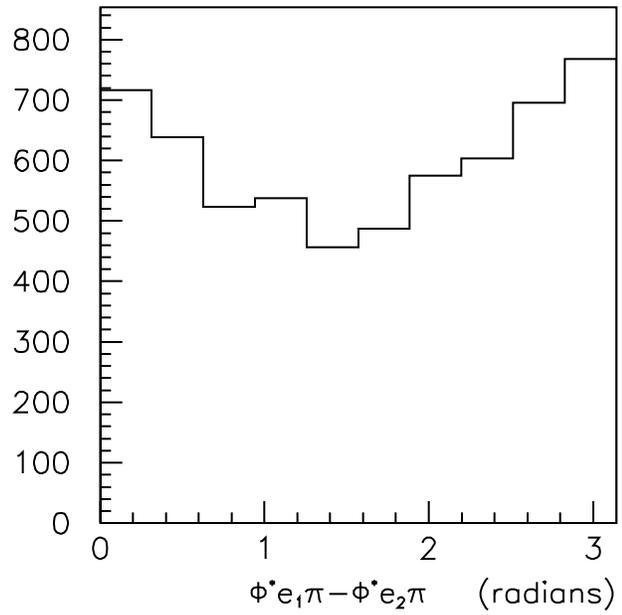
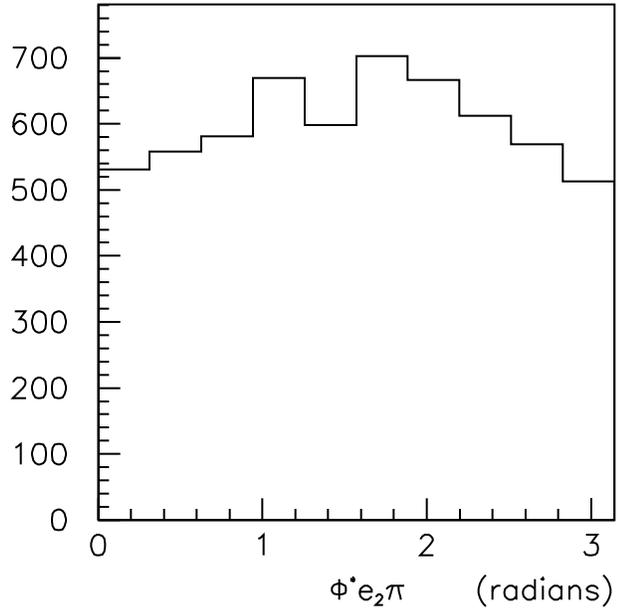
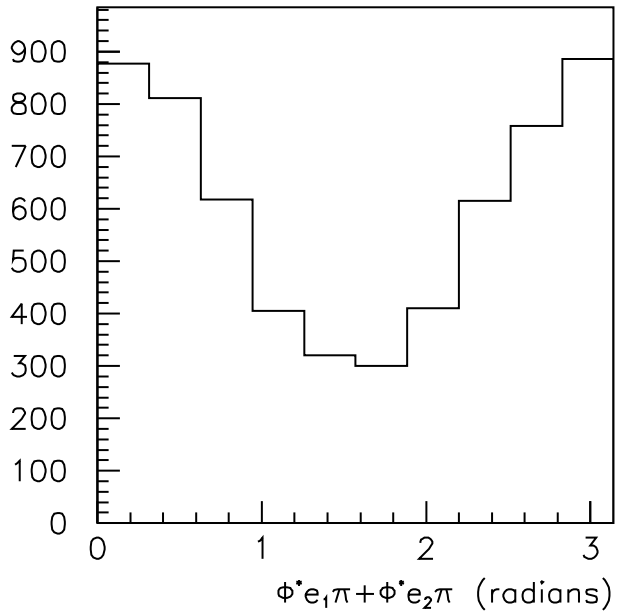
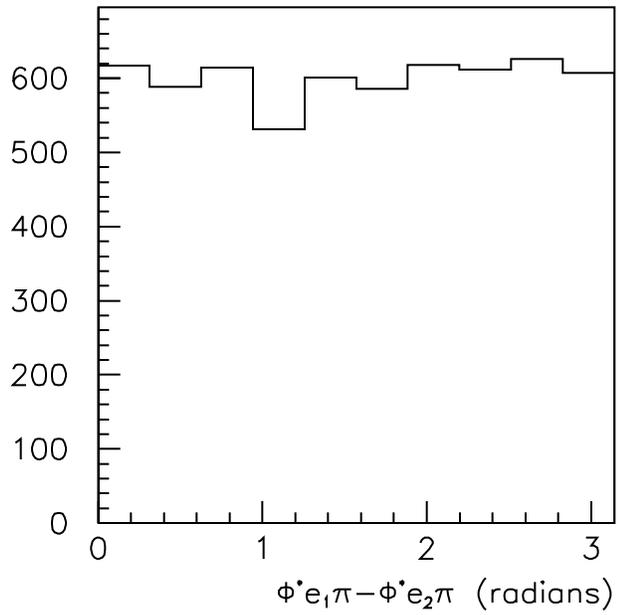
\end{document}